\documentclass[12pt]{article}
\usepackage{amssymb}

\usepackage{cite}

\def\be{\begin{equation}}
\def\ee{\end{equation}}
\def\bea{\begin{eqnarray}}
\def\eea{\end{eqnarray}}
\def\beb{\begin{eqnarray*}}
\def\eeb{\end{eqnarray*}}
\def\pat{\partial}

\begin{document}
\makeatletter
\def\fmslash{\@ifnextchar[{\fmsl@sh}{\fmsl@sh[0mu]}}
\def\fmsl@sh[#1]#2{%
  \mathchoice
    {\@fmsl@sh\displaystyle{#1}{#2}}%
    {\@fmsl@sh\textstyle{#1}{#2}}%
    {\@fmsl@sh\scriptstyle{#1}{#2}}%
    {\@fmsl@sh\scriptscriptstyle{#1}{#2}}}
\def\@fmsl@sh#1#2#3{\m@th\ooalign{$\hfil#1\mkern#2/\hfil$\crcr$#1#3$}}
\makeatother
\thispagestyle{empty}
\begin{titlepage}

\hspace*{\fill}{{\normalsize \begin{tabular}{l}
                              {\sf UWThPh-2005-9}\\
                              \end{tabular} }}                                                                                                                                  
                                                                                                                                         
\boldmath
\begin{center}
  {\Large {\bf On $\kappa$-Deformation and UV/IR Mixing}}
\end{center}
\unboldmath

\setcounter{footnote}{0}
\renewcommand{\thefootnote}{\arabic{footnote}}
\begin{center}
                                                                                                                                         
{{{\bf
Harald~Grosse\footnote{email: harald.grosse@univie.ac.at}
and
Michael~Wohlgenannt\footnote{email: michael.wohlgenannt@univie.ac.at}
}}}
                                                                                                                                         
\end{center}
\vskip 1em
\begin{center}
Universit\"at Wien, Institut f\"ur Theoretische Physik, \\
  Boltzmanngasse 5, 1090 Wien, Austria\\
                                                                                                                                         
\end{center}

\vspace{\fill}
                                                                                                                                         
\begin{abstract}
\noindent
We examine the UV/IR mixing property on a $\kappa$-deformed Euclidean space for a real
scalar $\phi^4$ theory. All contributions
to the tadpole diagram are explicitly calculated. UV/IR mixing is present, though in a different
dressing than in the case of the canonical deformation.
\end{abstract}
\begin{center}
\end{center}

\end{titlepage}


\section{Introduction}

Divergencies in quantum field theory have been one of the main reasons for introducing
non-commutative geometries and non-commutative coordinates. In the simplest case, the commutator of two coordinates is 
just constant,
$$
{}[\hat x^\mu, \hat x^\nu] = i\theta^{\mu\nu},
$$
where $\theta^{\mu\nu}\in \mathbb R$.
This is the canonical deformation. There is an enormous amount of literature dealing with field theories built on
such spaces. Feynman rules for scalar $\phi^4$ theory have been deduced in
\cite{Filk:1996dm,Denk:2003jj}. Unfortunately, non-commutative field theories turned out to be non-renormalisable
due to a new property, called "UV/IR mixing" \cite{Minwalla:1999px}. Although the one-loop integrals for
non-planar diagrams are finite for generic external momenta, it diverges for zero external momenta. This causes
infrared problems even in massive theories. The insertion into higher-loop contributions gives rise to 
divergencies which cannot be absorbed by standard procedures.

So far, the only renormalisable model on non-commutative spaces has been provided by R.~Wulkenhaar and one
of the authors (H.G.) \cite{Grosse:2003nw,Grosse:2004yu}.

In this work, we want to study the UV/IR mixing property for real scalar $\phi^4$ theory on $\kappa$-deformed
Euclidean space. In the $\kappa$-deformed case \cite{Lukierski:1991pn, Majid:1994cy, Dimitrijevic:2003wv,
Dimitrijevic:2004nv, Dimitrijevic:2004vv}, the space algebra is spanned by coordinates 
$\hat x^i$, $i=1,2,3,4$ with relations
\be
[ \hat x^1, \hat x^p] = i a \hat x^p, \qquad
[\hat x^q, \hat x^p] =0,
\ee
where $p,q=2,3,4$, $a=1/\kappa$. Elements of the non-commutative space-time algebra and
abstract elements of the $\kappa$-Poincar\'e algebra $\mathcal U_a(so(4))$ are denoted with a hat.

The Klein-Gordon operator is given by \cite{Dimitrijevic:2003wv}
\be
\widehat \Box = e^{-i a \hat \pat_1} \, \sum_{p=2}^4\hat \pat_p \hat \pat_p + {2 \over a^2} (1-\cos(a\hat\pat_1)).
\ee
In the following, summation over repeated indices is implied.
The algebra relations of the symmetry generators are given in \cite{Dimitrijevic:2004nv},
for example. The action of the generators on
commutative functions - star product representations - are provided in \cite{Dimitrijevic:2004vv},
with respect to various different orderings.
The ordering defines a basis in the abstract coordinate
algebra and therefore a star product representation on commutative functions.
The ordering is
not essential, though. The different star products corresponding to different ordering
prescriptions are equivalent and related by a 
(gauge) transformation $\mathcal D$ \cite{Jurco:2001my},
$$
\mathcal D f * \mathcal D g = \mathcal D (f*'g).
$$
Physics should only depend on equivalence classes of star products, not on the
representations of a single class.
We will concentrate on the symmetrically ordered star product. Its advantage is
the hermiticity property,
\be
\overline{f*g\,(x)} = \bar g * \bar f \, (x). 
\ee

Scalar field theories on $\kappa$-deformed spaces have already been studied in e.g.
\cite{Kosinski:1999ix,Amelino-Camelia:2001fd, Amelino-Camelia:2002mu,Daszkiewicz:2004xy, Nowak:2005qa}. In 
\cite{Amelino-Camelia:2001fd, Amelino-Camelia:2002mu}, a functional approach has been applied which we will also
adopt here. Nevertheless, no explicit results for Feynman amplitudes have been calculated yet.
A main difficulty was the construction of a proper measure \cite{Daszkiewicz:2004xy, Nowak:2005qa, 
Moller:2004sk, Agostini:2004cu}. For our calculations, we will choose a symmetrically ordered star product and
the $\kappa$-Poincar\'e invariant scalar product introduced in \cite{Moller:2004sk}.


\section{Symmetrically Ordered Star Product}

\noindent
The symmetrically ordered star product is given by \cite{Dimitrijevic:2004vv}
\bea
\label{star}
f*g\, (x) = \int d^4k \, d^4 p \, \tilde f(k) \tilde g(p)
 \, e^{i(\omega_k + \omega_p)x^1} e^{i\vec x( \vec k e^{a\omega_p} A(\omega_k,\omega_p)+ \vec p A(\omega_p,\omega_k))},
\eea
where $k=(\omega_k,\vec k)$, and $\vec x=(x^2,x^3,x^4)$.
We have used the definition
\be
\label{A}
A(\omega_k,\omega_p) \equiv  \frac{a(\omega_k+\omega_p)}{e^{a(\omega_k+\omega_p)}-1} 
	\frac{e^{a\omega_k}-1}{a\omega_k}.
\ee
Let us state a very useful identity which we will need a lot in  the calculations:
\be
\label{id1}
e^{-a\omega_2} A(-\omega_1,-\omega_2) = A(\omega_1,\omega_2).
\ee
Then, the Klein-Gordon operator acting on commutative functions reads
\be
\Box^* = \sum_{i=1}^4 \pat_i\pat_i \frac{2(1-\cos a\pat_1)}
	{a^2\pat_1^2}.
\ee

\subsection{$\kappa-$Poincar\'e Invariant Action}

A $\kappa$-Poincar\'e invariant integral is given by \cite{Moller:2004sk}
\be
(\phi, \psi) = \int d^4 x \phi ( K \bar\psi),
\ee
where 
\be
K = \left( \frac{ -ia\pat_1}{e^{-ia\pat_1}-1} \right)^{3}.
\ee
In momentum space, this amounts to 
\be
(\phi,\psi) = \int d^4 q \left( \frac{-a\omega_q}{e^{-a\omega_q}-1}\right)^3 \, \tilde \phi(q) \bar{\tilde\psi}(q).
\ee
And therefore, the action for a scalar field with $\phi^4$ interaction is given by
\bea
\label{dint}
S[\phi] & = & -(\phi, (\Box^* -m^2)\psi) \\
\nonumber
&& +\,  \frac{g}{4!} \left( b(\, \phi*\phi, \, \phi*\phi)
+ d (\,\phi*\phi*\phi*\phi,1)
\right).
\eea
In momentum space, the action has the following form:
\bea
\label{action}
S[\phi] & = &  \int d^4q \left( \frac{-a\omega_q}{e^{-a\omega_q}-1}\right)^3
\tilde \phi(q) \left(
q^2 \frac{2(\cosh a\omega_q - 1)}{a^2 \omega_q^2} + m^2
\right)
\bar{\tilde\phi}(q)
\\
\nonumber
&& 
\hspace{-.2cm}
+ b\,\frac{g}{4!} \int d^4 z  \prod_{i=1}^4 d^4 k_i
\left( \frac{a(\omega_{k_3}+\omega_{k_4})}{e^{a(\omega_{k_3}+\omega_{k_4})}-1} \right)^3
\, 
\tilde\phi(k_1)\tilde\phi(k_2) \tilde\phi(k_3)\tilde\phi(k_4) \\
\nonumber
&& \times
	e^{i z^1 \sum \omega_{k_i}}\,
	exp\bigg(i\vec z \left[
	\vec k_1 e^{a\omega_{k_2}} A(\omega_{k_1},\omega_{k_2}) + 
	\vec k_2 A(\omega_{k_2},\omega_{k_1})
\right.
\\ 
\nonumber && \hspace{.5cm} \left.+
	\vec k_3 e^{-a\omega_{k_4}} A(-\omega_{k_3},-\omega_{k_4}) +
	\vec k_4 A(-\omega_{k_4},-\omega_{k_3})
	\right] \bigg)
\nonumber
\\
\nonumber
&& 
\hspace{-.2cm}
+ d\,\frac{g}{4!} \int d^4 z  \prod_{i=1}^4 d^4 k_i
\, e^{i z^1 \sum \omega_{k_i}}
\tilde\phi(k_1)\tilde\phi(k_2) \tilde\phi(k_3)\tilde\phi(k_4) \\
\nonumber
&& \times
	e^{i\vec z 
	( \vec k_1 e^{a\omega_{k_2}}  A(\omega_{k_1},\omega_{k_2}) 
		+ \vec k_2 A( \omega_{k_2}, \omega_{k_1}) ) e^{a(\omega_{k_3}+\omega_{k_4})} 
		A(\omega_{k_1}+\omega_{k_2},\omega_{k_3}+\omega_{k_4}) }
		\\
\nonumber
&& \times
	e^{i \vec z ( \vec k_3 e^{a\omega_{k_4}}  A(\omega_{k_3},\omega_{k_4}) 
		+ \vec k_4 A( \omega_{k_4}, \omega_{k_3}) ) A(\omega_{k_3}+\omega_{k_4},\omega_{k_1}+\omega_{k_2})
	}.
\eea
Note that $ \bar{\tilde \phi}(k) = \tilde \phi(-k)$, for real fields $\phi(x)$. The $x$-dependent
phase factors are a direct result of the star product (\ref{star}), $b$ and $d$ are real parameters. In the 
case of canonical deformation, the phase factor is independent of $x$. 
In (\ref{dint}), one could also imagine an interaction term proportional to $(\phi*\phi*\phi,\phi)$.
This term will be discussed later on. It leads to a somewhat different and peculiar behaviour.


\section{Tadpole Diagram}

The generating functional can be defined as 
\be
\label{f1}
Z_\kappa[J] = \int \mathcal D\phi e^{-S[\phi] + \frac12 (J,\phi) + \frac12 (\phi,J) }.
\ee
The $n$-point functions $\tilde G_n(p_1,\dots,p_n)$ are given by functional
differentiation:
\be
\label{f2}
\tilde G_n(p_1,\dots,p_n) = \frac{\delta^n}{\delta \tilde J(-p_1) \dots \delta
\tilde J (-p_n) } Z_\kappa [J]\Big|_{J=0}.
\ee
Let us first consider the free case. For the free generating functional
$Z_{0,\kappa}$ we obtain from Eq.~(\ref{f1})
\bea
\nonumber
Z_{0,\kappa}[J] & = & \int \mathcal D\phi \exp \left[
-\frac12 \int d^4k \left( \frac{-a \omega_k}{e^{-a\omega_k}-1} \right)^3
\tilde\phi (k) ( \mathcal M_k + m^2 ) \tilde \phi(-k)
	\right.\\
\label{f3}&& \hspace{-.5cm}
\left.  +\frac12 \int d^4k \left( \left( \frac{-a \omega_k}{e^{-a\omega_k}-1}
\right)^3+ \left( \frac{a \omega_k}{e^{a\omega_k}-1} \right)^3 \right) \tilde
J(k) \, \tilde \phi(-k)
	\right],
\eea
where we have defined
\be
\mathcal M_k := \frac{ 2 k^2 (\cosh a\omega_k -1)}{a^2 \omega_k^2}.
\ee
The same manipulations as in the classical case yield
\be
\label{f4}
Z_{0,\kappa}[J] =Z_{0,\kappa}[0] e^{ \frac12 \int d^4k 
\left( \frac{-a \omega_k}{e^{-a\omega_k}-1} \right)^3 \frac{\tilde J(k) 
\tilde J(-k)}{\mathcal M_k+m^2}}.
\ee
We will always consider the normalised functional, which we obtain by dividing
with $Z_{0,\kappa}[0]$.
Now, the free propagator is given by
\bea
\label{f5}
\tilde G(k,p) & = & \frac{\delta^2}{\delta \tilde J(-k) \delta \tilde J(-p)}
Z_{0,\kappa}[J]\Big|_{J=0}\\
\nonumber & = & 
L(\omega_k) \frac{\delta^{(4)} (k+p)}{\mathcal M_k + m^2}\equiv \delta^{(4)}(k+p)Q_k. 
\eea
For the sake of brevity, we have introduced
\be
\label{f6}
L(\omega_k) := \frac12 \left( 
\left( \frac{-a \omega_k}{e^{-a\omega_k}-1} \right)^3 +
\left( \frac{a \omega_k}{e^{a\omega_k}-1} \right)^3 \right).
\ee

Let us switch on the interaction. For the clarity of presentation, we will for now
only consider the first interaction term in Eq.~(\ref{action}). The other term
will be treated in the next subsection. We make the following observation:
\be
\frac1{L(\omega_p)} \, \frac {\delta}{\delta \tilde J(-p)} Z_\kappa[J]\Big|_{J=0} = \tilde \phi
(p).
\ee
Therefore, we can rewrite the generating functional in the form
\be
\label{f7}
Z_\kappa[J] = e^{-S_I[ 1/L(\omega_k) \frac{\delta}{\delta \tilde J(-k)}]}
Z_{0,\kappa}[J].
\ee
The aim of this article is to compute tadpole diagram contributions. In order
to do so, we expand the generating functional (\ref{f7}) in powers of the
coupling constant $g$. Using Eq.~(\ref{action}), we obtain
\be
\label{f8}
Z_\kappa[J] = Z_{0,\kappa}[J] + Z_\kappa^1[J] + \mathcal O(g^2).
\ee
The first order term in this expansion reads
\bea
\nonumber
Z_\kappa^1[J] & = & - b \frac{g}4 \int \prod_{i=1}^4  \left ( d^4k_i \frac{
	\left( \frac{a(\omega_3+\omega_4)}{e^{a(\omega_3+\omega_4)}-1}\right)^3
	}{L(\omega_i)} \frac{\delta}{\delta \tilde J(-k_i)}
	\right) Z_{0,\kappa}[J]
	 \\
\nonumber 
&& 
\times 
\delta(\sum_{j=1}^4 \, \omega_j) \delta^{(3)} \Big( \vec k_1 e^{a\omega_2}
A(\omega_1,\omega_2) + \vec k_2 A(\omega_2,\omega_1) \\
\nonumber && \hspace{1.5cm} + \vec k_3 e^{-a\omega_4}
A(-\omega_3,-\omega_4) + \vec k_4 A(-\omega_4,\omega_3)
	\Big)\\
\label{f10}
& = &
- b \frac{g}4 \int \prod_{i=1}^4 \left( d^4k_i \frac{
	\left( \frac{a(\omega_3+\omega_4)}{e^{a(\omega_3+\omega_4)}-1}\right)^3
	}{L(\omega_i)} \right) \, \delta(\sum_{j=1}^4 \omega_j)\\
\nonumber && 
\times \delta^{(3)} \Big( \vec k_1 e^{a\omega_2}
A(\omega_1,\omega_2) + \vec k_2 A(\omega_2,\omega_1)\\
\nonumber && \hspace{1.5cm} 
 + \vec k_3 e^{-a\omega_4}
A(-\omega_3,-\omega_4) + \vec k_4 A(-\omega_4,-\omega_3)
	\Big)\\
\nonumber && \hspace{-.2cm} \times \Big\{
\delta^{(4)}(k_1+k_2) \delta^{(4)}(k_3+k_4) Q_4 Q_2\\
\nonumber && 
+ \delta^{(4)}(k_1+k_3) \delta^{(4)}(k_2+k_4) Q_4 Q_3
\\
\nonumber && + 
\delta^{(4)}(k_1+k_4) \delta^{(4)}(k_2+k_3) Q_4 Q_3 \\
\nonumber && + 
Q_4 \tilde J(k_4) Q_3 \tilde J(k_3) Q_2 \tilde J(k_2) Q_1 \tilde J(k_1) 
\\
\nonumber && +
\delta^{(4)}(k_3+k_4) Q_4 Q_1 \tilde J(k_1) Q_2 \tilde J(k_2) \\
\nonumber && +
\delta^{(4)}(k_2+k_4) Q_4 Q_1 \tilde J(k_1) Q_3 \tilde J(k_3)\\
\nonumber && +
\delta^{(4)}(k_2+k_3) Q_3 Q_1 \tilde J(k_1) Q_4 \tilde J(k_4)\\
\nonumber && +
\delta^{(4)}(k_1+k_4) Q_4 Q_2 \tilde J(k_2) Q_3 \tilde J(k_3)\\
\nonumber && +
 \delta^{(4)}(k_1+k_3) Q_3 Q_2 \tilde J(k_2) Q_4 \tilde J(k_4)\\
\nonumber && +
 \delta^{(4)}(k_1+k_2) Q_2 Q_3 \tilde J(k_3) Q_4 \tilde J(k_4)
\Big\} Z_{0,\kappa}[J],
\eea
where $Q_i:=\frac{L(\omega_i)}{\mathcal M_{k_i} + m^2}$, and $k_j=(\omega_j, \vec k_j)$.
The full propagator to first order in the coupling parameter is given by the
connected part of the expression
\be
\label{f11}
\tilde G^{(2)}(p,q) = \frac{\delta^2}{\delta \tilde J(-p) \delta \tilde J(-q)}
Z_\kappa[J]\Big|_{J=0}. 
\ee
The first three terms of Formula~(\ref{f10}) give
the disconnected contribution to the 2-point function. The contribution of the
fourth term vanishes. What remains, provides us with twelve contributions to the
connected 2-point function. Explicitly, we obtain
\bea
\label{f12}
\tilde G^{(2)}(p,q) & = & \tilde G(p,q) - b \frac{g}{4!} \int_\Lambda \prod_{i=1}^4 \left( d^4k_i
\frac{\left( \frac{a(\omega_3+\omega_4)}{e^{a(\omega_3+\omega_4)}-1}\right)^3
}{L(\omega_i)} \right)
\\
\nonumber && \hspace{-1cm} \times
\delta(\sum _{j=1}^4 \omega_j) 
\, \delta^{(3)} \Big( \vec k_1 e^{a\omega_2} A(\omega_1,\omega_2)
	+ \vec k_2 A (\omega_2,\omega_1) 
\\
\nonumber &&  \hspace{1cm}
+ \vec k_3 e^{-a\omega_4} A(-\omega_3,-\omega_4) + \vec k_4 A(-\omega_4,-\omega_3)
\Big)\\
\nonumber && \hspace{-.2cm} \times \left\{
\delta^{(4)}(k_2+p) \delta^{(4)}(k_1+q) \delta^{(4)}(k_3+k_4) Q_4 Q_2 Q_1 
\right.\\
\nonumber && \,\, + 
\delta^{(4)}(k_1+p) \delta^{(4)}(k_2+q) \delta^{(4)}(k_3+k_4) Q_4 Q_2 Q_1
\\
\nonumber && \,\, + 
\delta^{(4)}(k_1+p) \delta^{(4)}(k_3+q) \delta^{(4)}(k_2+k_4) Q_4 Q_3 Q_1 
\\
\nonumber && \,\, + 
\delta^{(4)}(k_3+p) \delta^{(4)}(k_1+q) \delta^{(4)}(k_2+k_4) Q_4 Q_3 Q_1 
\\
\nonumber && \,\, + 
\delta^{(4)}(k_4+p) \delta^{(4)}(k_2+q) \delta^{(4)}(k_1+k_3) Q_4 Q_3 Q_2 
\\
\nonumber && \,\, + 
\delta^{(4)}(k_2+p) \delta^{(4)}(k_4+q) \delta^{(4)}(k_1+k_3) Q_4 Q_3 Q_2 
\\
\nonumber && \,\, + 
\delta^{(4)}(k_3+p) \delta^{(4)}(k_4+q) \delta^{(4)}(k_1+k_2) Q_4 Q_3 Q_2
\\
\nonumber && \,\, + 
\delta^{(4)}(k_4+p) \delta^{(4)}(k_3+q) \delta^{(4)}(k_1+k_2) Q_4 Q_3 Q_2
\\
\nonumber && \,\, + 
\delta^{(4)}(k_4+p) \delta^{(4)}(k_1+q) \delta^{(4)}(k_2+k_3) Q_4 Q_3 Q_1 
\\
\nonumber && \,\, + 
\delta^{(4)}(k_1+p) \delta^{(4)}(k_4+q) \delta^{(4)}(k_2+k_3) Q_4 Q_3 Q_1 
\\
\nonumber && \,\, + 
\delta^{(4)}(k_3+p) \delta^{(4)}(k_2+q) \delta^{(4)}(k_1+k_4) Q_4 Q_3 Q_2 
\\
\nonumber && \left. \,\, + 
\delta^{(4)}(k_2+p) \delta^{(4)}(k_3+q) \delta^{(4)}(k_1+k_4) Q_4 Q_3 Q_2 
\right\}\\
\label{f12a}
& \equiv & \tilde G(p,q) + b \sum_{i=1}^{12} \tilde G^{(2),b}_{c=i}(p,q),
\eea
where $\int_\Lambda$ denotes the integral regularised by a cut-off $\Lambda$, see below.
The last four terms of Eq.~(\ref{f12}) correspond to non-planar diagrams.
Let us discuss some of the contributions in detail. As an example of a planar
diagram, we will first of all analyse the first term of formula~(\ref{f12}), 
i.e., $c=1$. 
The $\delta$-functions from functional differentiation enable us to integrate 
over three of the momenta. We obtain the following contribution:
\bea
\nonumber
\tilde G^{(2),b}_{c=1}(p,q) & = &
-\frac{g}{4!} \int_\Lambda \prod_{i=1}^4 \left( d^4k_i \frac{\left( \frac{a(\omega_3+\omega_4)}{e^{a(\omega_3+\omega_4)}-1}\right)^3
}{L(\omega_i)}\right) \, \delta(\sum_j \omega_j) Q_4 Q_2 Q_1
\\
\nonumber && \hspace{-1cm}
\times \delta^{(3)} \Big( \vec k_1 e^{a\omega_2}
A(\omega_1,\omega_2) + \vec k_2 A(\omega_2,\omega_1) + \vec k_3 e^{-a\omega_4}
A(-\omega_3,-\omega_4) \\
\nonumber &&  \hspace{.5cm}
+ \vec k_4 A(-\omega_4,\omega_3) \Big) 
\delta^{(4)}(k_2+p) \delta^{(4)}(k_1+q) \delta^{(4)}(k_3+k_4)\\
\nonumber
& = & -\frac{g}{4!} \delta(\omega_q+\omega_p) \frac1{\mathcal M_p + m^2}
\frac1{\mathcal M_q + m^2}
\int_\Lambda d^4k \frac1{L(\omega_k)} \frac1{\mathcal M_k +
m^2}\\
\label{f13} &&  \times
\delta^{(3)}(\vec k(e^{a\omega_k} A(-\omega_k,\omega_k)-A(\omega_k,-\omega_k))
\\
\nonumber && \hspace{2cm}
-\vec qe^{a\omega_q}A(-\omega_q,\omega_q)-\vec p A(\omega_q,-\omega_q)) \\
\nonumber
& = & -\frac{g}{4!} \delta^{(4)}(p+q)\frac1{\mathcal M_p + m^2}
\frac1{\mathcal M_q + m^2}\frac1{|A(\omega_q,-\omega_q)|^3} 
\\
\nonumber && \hspace{1.5cm} \times
\int_\Lambda d^4 k
\frac1{L(\omega_k)} \frac1{\mathcal M_k+m^2}\\
\nonumber
& = & -\frac{g}{4!} \delta^{(4)}(p+q)\frac1{\mathcal M_p + m^2}
\frac1{\mathcal M_q + m^2}\frac1{|A(\omega_q,-\omega_q)|^3} \\
\nonumber && \hspace{-1.5cm}  \times
\int_{-\infty}^\infty d\omega_k\frac{4\pi}{L(\omega_k)}
\frac{a^2 \omega_k^2}{2(\cosh a\omega_k-1)} \int_0^\Lambda dk
\frac{k^2}{k^2+(\omega_k^2+\frac{m^2a^2\omega_k^2}{2(\cosh a\omega_k -1)})}\\
\label{f14}
& = & -\frac{g}{4!} \delta^{(4)}(p+q)\frac1{\mathcal M_p + m^2}
\frac1{\mathcal M_q + m^2}\frac1{|A(\omega_q,-\omega_q)|^3} \\
\nonumber &&  \hspace{-1cm} \times
\int_{-\infty}^\infty d\omega_k\frac{4\pi}{L(\omega_k)}
\frac{a^2 \omega_k^2}{2(\cosh a\omega_k-1)} 
\\
\nonumber && \hspace{-.5cm} \times
\left(
\Lambda - \omega_k\sqrt{ 1+\frac{m^2a^2}{2(\cosh a\omega_k-1)}} 
\arctan \frac{\Lambda}{ \omega_k\sqrt{ 1+\frac{m^2a^2}{2(\cosh a\omega_k-1)}}}
\right).
\eea
This expression is linearly divergent in the cut-off $\Lambda$. The $\omega_k$-integration
yields a finite result due to the propagator (\ref{f5}). For the other
planar diagrams, we obtain similiar results:
\bea
\label{f15}
\tilde G_{c=3}^{(2),b}(p,q) & = & 
-\frac{g}{4!} \delta^{(4)}(p+q)\frac1{\mathcal M_p + m^2}
\frac1{\mathcal M_q + m^2} \\
\nonumber &&  \hspace{-1.5cm} \times
\int_{-\infty}^\infty d\omega_k\frac{4\pi \left(\frac{-a(\omega_q+\omega_k)}
	{e^{-a(\omega_q+\omega_k)}-1}\right)^3}{L(\omega_k)}
\frac{a^2 \omega_k^2}{2(\cosh a\omega_k-1)} \frac1{|e^{a\omega_k}
	A(\omega_q,\omega_k)|^3}
\\
\nonumber && \hspace{-1cm} \times
\left(
\Lambda - \omega_k\sqrt{ 1+\frac{m^2a^2}{2(\cosh a\omega_k-1)}} 
\arctan \frac{\Lambda}{ \omega_k\sqrt{ 1+\frac{m^2a^2}{2(\cosh a\omega_k-1)}}}
\right),\\
\label{f16}
\tilde G_{c=5}^{(2),b}(p,q) & = & 
-\frac{g}{4!} \delta^{(4)}(p+q)\frac1{\mathcal M_p + m^2}
\frac1{\mathcal M_q + m^2} \\
\nonumber &&  \hspace{-1.5cm} \times
\int_{-\infty}^\infty d\omega_k\frac{4\pi \left(\frac{a(\omega_q+\omega_k)}
	{e^{a(\omega_q+\omega_k)}-1}\right)^3}{L(\omega_k)}
\frac{a^2 \omega_k^2}{2(\cosh a\omega_k-1)} \frac1{|e^{a\omega_k}
	A(\omega_q,\omega_k)|^3}
\\
\nonumber && \hspace{-1cm} \times
\left(
\Lambda - \omega_k\sqrt{ 1+\frac{m^2a^2}{2(\cosh a\omega_k-1)}} 
\arctan \frac{\Lambda}{ \omega_k\sqrt{ 1+\frac{m^2a^2}{2(\cosh a\omega_k-1)}}}
\right),\\
\label{f17}
\tilde G_{c=7}^{(2),b}(p,q) & = & 
-\frac{g}{4!} \delta^{(4)}(p+q)\frac1{\mathcal M_p + m^2}
\frac1{\mathcal M_q + m^2}\frac1{|A(\omega_q,-\omega_q)|^3} \\
\nonumber && \hspace{-1.5cm} \times
\int_{-\infty}^\infty d\omega_k\frac{4\pi}{L(\omega_k)}
\frac{a^2 \omega_k^2}{2(\cosh a\omega_k-1)} 
\\
\nonumber && \hspace{-1cm} \times
\left(
\Lambda - \omega_k\sqrt{ 1+\frac{m^2a^2}{2(\cosh a\omega_k-1)}} 
\arctan \frac{\Lambda}{ \omega_k\sqrt{ 1+\frac{m^2a^2}{2(\cosh a\omega_k-1)}}}
\right).
\eea
The remaining planar contractions $c=2,4,6,8$ can be obtained from the
contributions of $c=1,3,5$ and $7$, respectively by interchanging the external
momenta $p$ and $q$.

Non-planar contributions show a remarkable difference. There we do not have an
overall momentum conservation. Remarkably, the components $\vec k$ of the
internal momentum are fixed by the external ones, and the contributions are
finite, for generic external momenta. The only exception is the case
$\omega_p=\omega_q=0$. There, we get back the UV divergences discussed above.  
Below, we give the explicit calculation for the non-planar contraction $c=9$:
\bea
\label{f18}
\nonumber
\tilde G^{(2),b}_{c=9}(p,q) & = &
-\frac{g}{4!} \int \prod_{i=1}^4 \left( d^4k_i \frac{
	\left( \frac{a(\omega_3+\omega_4)}{e^{a(\omega_3+\omega_4)}-1}\right)^3
	}{L(\omega_i)} \right) 
	\delta(\sum_j \omega_j) Q_4 Q_3 Q_1
\\
\nonumber && \hspace{-1cm}
\times \delta^{(3)} \Big( \vec k_1 e^{a\omega_2}
A(\omega_1,\omega_2) + \vec k_2 A(\omega_2,\omega_1) + \vec k_3 e^{-a\omega_4}
A(-\omega_3,-\omega_4) \\
\nonumber && \hspace{.5cm} 
+ \vec k_4 A(-\omega_4,\omega_3) \Big)
 \delta^{(4)}(k_4+p) \delta^{(4)}(k_1+q) \delta^{(4)}(k_2+k_3)\\
\nonumber
& = & -\frac{g}{4!} \delta(\omega_q+\omega_p) \frac1{\mathcal M_p + m^2}
\frac1{\mathcal M_q + m^2} \int d^4k 
	\frac{\left(\frac{a(\omega_k+\omega_q)}{e^{a(\omega_q+\omega_k)}-1}
	\right)^3}{L(\omega_k)} \\
\label{f19} &&  \times \frac1{\mathcal M_k + m^2}\, 
\delta^{(3)}(\vec k(e^{-a\omega_q} A(-\omega_k,-\omega_q)-A(-\omega_k,-\omega_q))
\\
\nonumber && \hspace{2.5cm} 
-\vec p A(-\omega_q,-\omega_k)-\vec q e^{-a\omega_k} A(-\omega_q,-\omega_k)) \\
\nonumber
& = & -\frac{g}{4!} \delta(\omega_q+\omega_p) \frac1{\mathcal M_p + m^2}
\frac1{\mathcal M_q + m^2}\\
\nonumber && \hspace{-1cm} \times
\int_{-\infty}^\infty d\omega_k 
	\frac{\left(\frac{a(\omega_k+\omega_q)}{e^{a(\omega_q+\omega_k)}-1}
	\right)^3}{L(\omega_k)} 
	\frac{a^2\omega_k^2}{2(\cosh a\omega_k-1)}
	\frac1{| A(\omega_k,\omega_q) (1-e^{a\omega_q}) |^3}\\
\nonumber &&  \hspace{-.5cm} \times
 	\Bigg[\omega_k^2 + \frac{(\vec q + \vec p e^{a\omega_k})^2
	A(\omega_q,\omega_k)^2}{A(\omega_k,\omega_q)^2(1-e^{a\omega_q})^2}+
\frac{m^2a^2\omega_k^2}{2(\cosh a\omega_k-1)} \Bigg]^{-1}\\
\label{f20}
& = & -\frac{g}{4!} \delta(\omega_q+\omega_p) \frac1{\mathcal M_p + m^2}
\frac1{\mathcal M_q + m^2}\\
\nonumber && \hspace{-1cm} \times
\int_{-\infty}^\infty d\omega_k 
	\frac{\left(\frac{a(\omega_k+\omega_q)}{e^{a(\omega_q+\omega_k)}-1}
	\right)^3}{L(\omega_k)} 
	\frac{a^2\omega_k^2}{2(\cosh a\omega_k-1)}
	\frac1{| A(\omega_k,\omega_q) (1-e^{a\omega_q}) |^3}\\
\nonumber && \hspace{-.5cm}\times	
\Bigg[\omega_k^2 + \frac{(\vec q + \vec p e^{a\omega_k})^2
a^2 \omega_k^2}{a^2\omega_q^2(1-e^{a\omega_k})^2}+
\frac{m^2a^2\omega_k^2}{2(\cosh a\omega_k-1)} \Bigg]^{-1}.
\eea
The above formula is true for generic momenta $p$ and $q$. Expression~(\ref{f20}) is finite
except for $\omega_q=\omega_p=0$. In this case, the $\delta^{(3)}$-function in
Eq.~(\ref{f19}) does not depend on the internal momentum $\vec k$, and we encounter the same
UV singularity as in the planar case. What remains of the $\delta^{(3)}$-distribution,
$$
\frac1{|A(0,\omega_k)|^3} \delta^{(3)} ( \vec q + \vec p e^{a\omega_k} ),
$$
gives a contribution only if $\vec p$ and $\vec q$ are parallel to each other.

The non-planar amplitude corresponding to the contraction $c=11$ reads
\bea
\nonumber
\tilde G^{(2),b}_{c=11}(p,q) & = &
-\frac{g}{4!} \delta(\omega_q+\omega_p) \frac1{\mathcal M_p + m^2}
\frac1{\mathcal M_q + m^2} \int d^4k 
	\frac{\left(\frac{a(\omega_k+\omega_q)}{e^{a(\omega_q+\omega_k)}-1}
	\right)^3}{L(\omega_k)} 
	\\
\nonumber &&  \hspace{-1cm} \times \frac1{\mathcal M_k + m^2}
\delta^{(3)}( \vec k A(\omega_k,\omega_q)(e^{a\omega_q}-1)
-( \vec q e^{a\omega_k}+\vec p) A(\omega_q,\omega_k)) \\
\label{f21} 
& = &
-\frac{g}{4!} \delta(\omega_q+\omega_p) \frac1{\mathcal M_p + m^2}
\frac1{\mathcal M_q + m^2} 
\\
\nonumber && \hspace{-1cm} \times
\int_{-\infty}^\infty d\omega_k
	\frac{\left(\frac{a(\omega_k+\omega_q)}{e^{a(\omega_q+\omega_k)}-1}
	\right)^3}{L(\omega_k)} 
	\frac{a^2 \omega_k^2}{2(\cosh a\omega_k -1)} \frac1{ | (1-e^{a\omega_q})A(\omega_k,\omega_q) |^3 }
	\\
\nonumber
&&  \hspace{-.5cm} \times
\left[ \omega_k^2 + \frac{ (\vec q e^{a\omega_k}+\vec p)^2 A(\omega_q,\omega_k)^2}
	{(e^{a\omega_q}-1)^2 A(\omega_k,\omega_q)^2  }
	+ \frac{ m^2 a^2 \omega_k^2} {2(\cosh a\omega_k -1)}
\right]^{-1}.
\eea
For the exceptional situation $\omega_q=0$, we again do not obtain an overall momentum conservation, but 
a $\delta^{(3)}$-distribution fixing the $\omega_k$ component,
\be
\frac1{|A(0,\omega_k)|^3} \delta^{(3)}(\vec q e^{a\omega_k} + \vec p).
\ee
As before, the diagram shows a linear UV divergence. Assuming that $p=q=0$, exerts no influence
the divergencies, because the $\omega_k$-integration is exponentially damped by the modified propagator.

\subsection{Contributions from $S_I=\frac{g}{4!} (\phi*\phi*\phi*\phi,1)$}

The connected $2-$point function involving all interactions written down in Eq.~(\ref{dint})
is given by
\be
\tilde G^{(2)}(p,q) = \tilde G(p,q) + \sum_{i=1}^{12} \left( b \tilde G^{(2),b}_{c=i}(p,q)
+ d \tilde G^{(2),d}_{c=i}(p,q) \right) + \mathcal O(g^2).
\ee
The contributions $\tilde G^{(2),d}_{c=i}(p,q)$ for the second interaction term in Eq.~(\ref{dint})
are obtained in the same way as described in the previous subsection. Also, they display the same
characteristic behaviour:
\bea
\label{f22}
\tilde G^{(2),d}_{c=1}(p,q) & = &
-\frac{g}{4!} \delta^{(4)}(p+q)\frac1{\mathcal M_p + m^2}
\frac1{\mathcal M_q + m^2}\frac1{|A(\omega_q,-\omega_q)|^3} \\
\nonumber &&  \hspace{-2cm} \times
\int_{-\infty}^\infty d\omega_k\frac{4\pi}{L(\omega_k)}
\frac{a^2 \omega_k^2}{2(\cosh a\omega_k-1)} 
\\
\nonumber && \hspace{-1.5cm} \times
\left(
\Lambda - \omega_k\sqrt{ 1+\frac{m^2a^2}{2(\cosh a\omega_k-1)}} 
\arctan \frac{\Lambda}{ \omega_k\sqrt{ 1+\frac{m^2a^2}{2(\cosh a\omega_k-1)}}}
\right)
,
\\
\label{f23}
\tilde G^{(2),d}_{c=3}(p,q) & = & -\frac{g}{4!} \delta(\omega_q+\omega_p) \frac1{\mathcal M_p + m^2}
\frac1{\mathcal M_q + m^2}
\int_{-\infty}^\infty d\omega_k 
	\frac1{L(\omega_k)} 
	\\
\nonumber && \hspace{-2cm} \times 
	\frac{a^2\omega_k^2}{2(\cosh a\omega_k-1)} 
	\frac1{| A(-\omega_q - \omega_k, \omega_k+\omega_q) |^3}
	\frac1{|A(\omega_k,\omega_q)(1-e^{a\omega_q})|^3}\\
\nonumber && \hspace{-1.5cm} \times
	\left( \omega_k^2 + \frac{(\vec q + \vec p e^{a\omega_k})^2
	a^2 \omega_k^2}{a^2\omega_q^2(1-e^{a\omega_k})^2}+
\frac{m^2a^2\omega_k^2}{2(\cosh a\omega_k-1)} \right)^{-1}
,
\\
\label{f24}
\tilde G^{(2),d}_{c=5}(p,q) & = & -\frac{g}{4!} \delta(\omega_q+\omega_p) \frac1{\mathcal M_p + m^2}
\frac1{\mathcal M_q + m^2}
\int_{-\infty}^\infty d\omega_k 
	\frac1{L(\omega_k)} 
\\
\nonumber && \hspace{-2cm} \times  
	\frac{a^2\omega_k^2}{2(\cosh a\omega_k-1)}
	\frac1{|A(\omega_k+\omega_q,-\omega_k-\omega_q)|^3}
	\frac1{|A(\omega_k,\omega_q)(1-e^{a\omega_q})|^3}\\
\nonumber && \hspace{-1.5cm} \times
	\left( \omega_k^2 + \frac{(\vec q e^{a\omega_k}+ \vec p )^2
	a^2 \omega_k^2}{a^2\omega_q^2(1-e^{a\omega_k})^2}+
\frac{m^2a^2\omega_k^2}{2(\cosh a\omega_k-1)} \right)^{-1}
,
\\
\label{f25}
\tilde G_{c=7}^{(2),d}(p,q) & = & 
-\frac{g}{4!} \delta^{(4)}(p+q)\frac1{\mathcal M_p + m^2}
\frac1{\mathcal M_q + m^2}\frac1{|A(-\omega_q,\omega_q)|^3} \\
\nonumber && \hspace{-2cm} \times
\int_{-\infty}^\infty d\omega_k\frac{4\pi}{L(\omega_k)}
\frac{a^2 \omega_k^2}{2(\cosh a\omega_k-1)} 
\\
\nonumber && \hspace{-1.5cm} \times
\left(
\Lambda - \omega_k\sqrt{ 1+\frac{m^2a^2}{2(\cosh a\omega_k-1)}} 
\arctan \frac{\Lambda}{ \omega_k\sqrt{ 1+\frac{m^2a^2}{2(\cosh a\omega_k-1)}}}
\right)
,
\\
\label{f26}
\tilde G_{c=9}^{(2),d}(p,q) & = & 
-\frac{g}{4!} \delta^{(4)}(p+q)\frac1{\mathcal M_p + m^2}
\frac1{\mathcal M_q + m^2} \frac1{|A(\omega_q,-\omega_q)|^3}\\
\nonumber &&  \hspace{-2cm} \times
\int_{-\infty}^\infty d\omega_k\frac{4\pi }{L(\omega_k)}
\frac{a^2 \omega_k^2}{2(\cosh a\omega_k-1)} 
\\
\nonumber && \hspace{-1.5cm} \times
\left(
\Lambda - \omega_k\sqrt{ 1+\frac{m^2a^2}{2(\cosh a\omega_k-1)}} 
\arctan \frac{\Lambda}{ \omega_k\sqrt{ 1+\frac{m^2a^2}{2(\cosh a\omega_k-1)}}}
\right)
,
\\
\label{f27}
\tilde G_{c=11}^{(2),d}(p,q) & = & 
-\frac{g}{4!} \delta^{(4)}(p+q)\frac1{\mathcal M_p + m^2}
\frac1{\mathcal M_q + m^2} \\
\nonumber && \hspace{-2cm} \times
\int_{-\infty}^\infty d\omega_k\frac{4\pi}{L(\omega_k)}
\frac{a^2 \omega_k^2}{2(\cosh a\omega_k-1)} 
\frac1{ | e^{-a\omega_k} A(\omega_q, -\omega_q) |^3}
\\
\nonumber && \hspace{-1.5cm} \times
\left(
\Lambda - \omega_k\sqrt{ 1+\frac{m^2a^2}{2(\cosh a\omega_k-1)}} 
\arctan \frac{\Lambda}{ \omega_k\sqrt{ 1+\frac{m^2a^2}{2(\cosh a\omega_k-1)}}}
\right)
.
\eea
Due to the different arrangement of fields in the scalar product, planar and non-planar graphs 
have changed their "position" (with respect to the numbering in $c$).

\subsection*{Remark}

The possible interaction term $(\phi*\phi*\phi,\phi)$ leads to a somewhat different behaviour.
The contraction $c=1$, for example, is proportional to 
\bea
\label{remark1}
&& \int d^4k \, \frac1{L(\omega_k)} \frac1{\mathcal M_k + m^2} 
	\left(\frac{-a\omega_k}{e^{-a\omega_k}-1}\right)^3 \\
\nonumber
&& \hspace{1cm} \times 
\delta^{(3)} \left( \vec k - e^{a\omega_k} A(0,\omega_k) A(\omega_q,-\omega_q) 
(\vec q + \vec p) \right).
\eea
In the limit $a\to 0$, the integral in Eq.~(\ref{remark1}) reduces to 
$$ 
\int d \omega_k \frac1{\omega_k^2 + (\vec p+\vec q)^2 +m^2} = \frac{\pi}{\sqrt{(\vec p + \vec q)^2 + m^2}},
$$
contrary to the examples above where the $\delta^{(3)}$-distribution did not depend on 
$\vec k$ and therefore did not act as a regulator.

The contribution $c=7$, in order to give another example, is of the form
\be
\label{remark2}
\delta (\omega_p+\omega_q) \delta^{(3)} (\vec p) \int_\Lambda d^4 k\, \frac1{L(\omega_k)} \frac1{\mathcal M_k + m^2},
\ee
which is independent of $q$ and therefore peculiar.


\section{Conclusions}

Using a generating functional approach, we have deduced the Feynman rules for scalar $\phi^4$ theory on
$\kappa$-deformed Euclidean space. We have calculated the tadpole contributions explicitly. As in the canonically deformed
theory, we can distinguish between planar and non-planar diagrams. The planar diagrams (\ref{f14} - \ref{f17}) and
(\ref{f22}, \ref{f25} - \ref{f27}), respectively  display a linear UV divergence. The non-planar graphs (\ref{f20}, \ref{f21}) and 
(\ref{f23}, \ref{f24}), respectively are finite for generic external momenta $p,q$. In the exceptional case
$\omega_p=\omega_q=0$, however, the amplitudes also diverge linearly in the UV cut-off $\Lambda$. This is the form
of appearance of UV/IR mixing on $\kappa$-deformed spaces. Considering $\kappa$-Minkowski space-time, UV/IR mixing is
also expected to show up in a similar way.

So far, we have only discussed the massive case. For the massless case $m=0$, the divergencies have a richer structure.
The planar diagrams also show linear divergences in the cut-off $\Lambda$. The $\omega_k$-integration (\ref{f14}) is also finite.
The integrand in the massive case is peaked at $\omega_k=0$, whereas in the massless case it vanishes there. There are two peaks, 
one below and
one above $\omega_k=0$. This behaviour displays similarities to a phase transition. In the non-planar case, the generic contribution is
again finite. For $\omega_q=0$, the amplitude diverges as described in the massive case. But there is an additional exceptional
configuration, namely $\vec q = - \vec p$ and $\omega_q\ne 0$. In this case, the divergence structure of the integrand of the
$\omega_k$-integration (eg. (\ref{f20})) has to be studied in more detail.

The basic difference to the case of canonical deformation \cite{Filk:1996dm} is the appearance of 
$x$-dependent phase factors in Eq.~(\ref{action}). Similar $x$-dependent
phase factors already occurred in \cite{Robbins:2003ry},
where the UV/IR mixing has been discussed for two other Lie algebra deformations of space-time.
But there, no generalised symmetry is present. The star products $*_{RS}$ have an especially simple form,
$$
f *_{RS} g\, (x) = f(x)\cdot g(x) + \mbox{ total divergence.}  
$$
Using the usual integration yields unmodified propagators. The only modifications are in the interaction part
of the action. They obtain quadratic divergencies for the planar contributions. On the contrary, in our case
the necessary modifications of the propagator and of the free action change the 
divergence of the planar graphs (and of the non-planar ones for exceptional momenta).


\subsection*{Acknowledgement}

This work has been supported by FWF (Austrian Science Fund), project P16779-N02.


\begin{thebibliography}{10}

\bibitem{Filk:1996dm}
T.~Filk, ``Divergencies in a field theory on quantum space,'' {\em Phys. Lett.}
  {\bf B376} (1996)
53--58.

\bibitem{Denk:2003jj}
S.~Denk and M.~Schweda, ``Time ordered perturbation theory for non-local
  interactions: {A}pplications to {NCQFT},'' {\em JHEP} {\bf 09} (2003) 032,
\href{http://www.arXiv.org/abs/hep-th/0306101}{{\tt hep-th/0306101}}.

\bibitem{Minwalla:1999px}
S.~Minwalla, M.~Van~Raamsdonk, and N.~Seiberg, ``Noncommutative perturbative
  dynamics,'' {\em JHEP} {\bf 02} (2000) 020,
\href{http://www.arXiv.org/abs/hep-th/9912072}{{\tt hep-th/9912072}}.

\bibitem{Grosse:2003nw}
H.~Grosse and R.~Wulkenhaar, ``Renormalisation of phi**4 theory on
  noncommutative {R}**2 in the matrix base,'' {\em JHEP} {\bf 12} (2003) 019,
\href{http://www.arXiv.org/abs/hep-th/0307017}{{\tt hep-th/0307017}}.

\bibitem{Grosse:2004yu}
H.~Grosse and R.~Wulkenhaar, ``Renormalisation of {$\phi^4$} theory on
  noncommutative {$R^4$} in the matrix base,'' {\em Commun. Math. Phys.} {\bf
  256} (2005) 305--374,
\href{http://www.arXiv.org/abs/hep-th/0401128}{{\tt hep-th/0401128}}.

\bibitem{Lukierski:1991pn}
J.~Lukierski, H.~Ruegg, A.~Nowicki, and V.~N. Tolstoi, ``{q} deformation of
  {Poincar\'{e}} algebra,'' {\em Phys. Lett.} {\bf B264} (1991)
331--338.

\bibitem{Majid:1994cy}
S.~Majid and H.~Ruegg, ``Bicrossproduct structure of $\kappa$-{Poincar\'{e}}
  group and noncommutative geometry,'' {\em Phys. Lett.} {\bf B334} (1994)
  348--354,
\href{http://arXiv.org/abs/hep-th/9405107}{{\tt hep-th/9405107}}.

\bibitem{Dimitrijevic:2003wv}
M.~Dimitrijevi\'c, L.~Jonke, L.~M{\"o}ller, E.~Tsouchnika, J.~Wess, and
  M.~Wohlgenannt, ``Deformed {Field Theory} on $\kappa$-spacetime,'' {\em Eur.
  Phys. J.} {\bf C31} (2003) 129--138,
\href{http://www.arXiv.org/abs/hep-th/0307149}{{\tt hep-th/0307149}}.

\bibitem{Dimitrijevic:2004nv}
M.~Dimitrijevic, L.~Jonke, L.~M{\"o}ller, E.~Tsouchnika, J.~Wess, and
  M.~Wohlgenannt, ``Field theory on $\kappa$-spacetime,'' {\em Czech. J. Phys.}
  {\bf 54} (2004) 1243--1248,
\href{http://www.arXiv.org/abs/hep-th/0407187}{{\tt hep-th/0407187}}.

\bibitem{Dimitrijevic:2004vv}
M.~Dimitrijevi\'c, L.~M{\"o}ller, and E.~Tsouchnika, ``Derivatives, forms and
  vector fields on the $\kappa$-deformed {E}uclidean space,'' {\em J. Phys.}
  {\bf A37} (2004) 9749--9770,
\href{http://www.arXiv.org/abs/hep-th/0404224}{{\tt hep-th/0404224}}.

\bibitem{Jurco:2001my}
B.~Jur\v{c}o, P.~Schupp, and J.~Wess, ``Nonabelian noncommutative gauge theory
  via noncommutative extra dimensions,'' {\em Nucl. Phys.} {\bf B604} (2001)
  148--180,
\href{http://arXiv.org/abs/hep-th/0102129}{{\tt hep-th/0102129}}.

\bibitem{Kosinski:1999ix}
P.~Kosi\'{n}ski, J.~Lukierski, and P.~Ma{\'s}lanka, ``Local {D = 4} field
  theory on $\kappa$-deformed {Minkowski} space,'' {\em Phys. Rev.} {\bf D62}
  (2000) 025004,
\href{http://arXiv.org/abs/hep-th/9902037}{{\tt hep-th/9902037}}.

\bibitem{Amelino-Camelia:2001fd}
G.~Amelino-Camelia and M.~Arzano, ``Coproduct and star product in field
  theories on {Lie}-algebra non-commutative space-times,'' {\em Phys. Rev.}
  {\bf D65} (2002) 084044,
\href{http://arXiv.org/abs/hep-th/0105120}{{\tt hep-th/0105120}}.

\bibitem{Amelino-Camelia:2002mu}
G.~Amelino-Camelia, M.~Arzano, and L.~Doplicher, ``Field theories on canonical
  and {Lie}-algebra noncommutative spacetimes,''
\href{http://arXiv.org/abs/hep-th/0205047}{{\tt hep-th/0205047}}.

\bibitem{Daszkiewicz:2004xy}
M.~Daszkiewicz, K.~Imilkowska, J.~Kowalski-Glikman, and S.~Nowak, ``Scalar
  field theory on $\kappa$-{M}inkowski space-time and {Doubly Special
  R}elativity,''
\href{http://www.arXiv.org/abs/hep-th/0410058}{{\tt hep-th/0410058}}.

\bibitem{Nowak:2005qa}
S.~Nowak, ``Lorentz invariance of scalar field action on $\kappa$- {M}inkowski
  space-time,''
\href{http://www.arXiv.org/abs/hep-th/0501017}{{\tt hep-th/0501017}}.

\bibitem{Moller:2004sk}
L.~M{\"o}ller, ``A symmetry invariant integral on $\kappa$-deformed
  spacetime,''
\href{http://www.arXiv.org/abs/hep-th/0409128}{{\tt hep-th/0409128}}.

\bibitem{Agostini:2004cu}
A.~Agostini, G.~Amelino-Camelia, M.~Arzano, and F.~D'Andrea, ``Action
  functional for $\kappa$-{M}inkowski {Noncommutative S}pacetime,''
\href{http://www.arXiv.org/abs/hep-th/0407227}{{\tt hep-th/0407227}}.

\bibitem{Robbins:2003ry}
D.~Robbins and S.~Sethi, ``The {UV/IR} interplay in theories with space-time
  varying non-commutativity,'' {\em JHEP} {\bf 07} (2003) 034,
\href{http://www.arXiv.org/abs/hep-th/0306193}{{\tt hep-th/0306193}}.

\end{thebibliography}

\providecommand{\href}[2]{#2}\begingroup\raggedright\endgroup

\end{document}